\documentclass[conference,compsoc]{IEEEtran}

\ifCLASSOPTIONcompsoc
  \usepackage[nocompress]{cite}
\else
  \usepackage{cite}
\fi
\usepackage{amsmath}
\usepackage{amssymb}
\usepackage{amsthm}
\DeclareMathOperator*{\argmax}{arg\,max}

\usepackage{algorithm}

\usepackage{algorithmic}

\usepackage{graphicx}

\usepackage{url}

\begin{document}
%
\title{Clustered Hierarchical Entropy-Scaling Search of Astronomical and Biological Data}

\author{
\IEEEauthorblockN{Najib Ishaq}
\IEEEauthorblockA{Computer Science and Statistics\\
University of Rhode Island\\
Kingston, RI 02881\\
najib\_ishaq@zoho.com}
\and
\IEEEauthorblockN{George Student}
\IEEEauthorblockA{Computer Science and Statistics\\
University of Rhode Island\\
Kingston, RI 02881}
\and
\IEEEauthorblockN{Noah M. Daniels}
\IEEEauthorblockA{Computer Science and Statistics\\
University of Rhode Island\\
Kingston, RI 02881\\
noah\_daniels@uri.edu}}


%


\maketitle

\begin{abstract}
Both astronomy and biology are experiencing explosive growth of data, resulting in a ``big data'' problem that stands in the way of a ``big data'' opportunity for discovery. One common question asked of such data is that of approximate search ($\rho-$nearest neighbors search). We present a hierarchical search algorithm for such data sets that takes advantage of particular geometric properties apparent in both astronomical and biological data sets, namely the metric entropy and fractal dimensionality of the data. We present CHESS (Clustered Hierarchical Entropy-Scaling Search), a search tool with virtually no loss in specificity or sensitivity, demonstrating a $13.6\times$ speedup over linear search on the Sloan Digital Sky Survey's APOGEE data set and a $68\times$ speedup on the GreenGenes 16S metagenomic data set, as well as asymptotically fewer distance comparisons on APOGEE when compared to the FALCONN locality-sensitive hashing library. CHESS demonstrates an asymptotic complexity not directly dependent on data set size, and is in practice at least an order of magnitude faster than linear search by performing fewer distance comparisons. Unlike locality-sensitive hashing approaches, CHESS can work with any user-defined distance function. CHESS also allows for implicit data compression, which we demonstrate on the APOGEE data set. We also discuss an extension allowing for efficient k-nearest neighbors search.
\end{abstract}


%
\IEEEpeerreviewmaketitle

\section{Introduction}

Astronomers have learned much of what they know about celestial objects by studying their electromagnetic spectra. By attaching an instrument known as a spectrograph to a telescope, astronomers are able to separate the incoming light by its wavelength, generating a fingerprint of the object being observed. We call this fingerprint a spectrum. Spectra can reveal information about characteristics of celestial objects including chemical composition, luminosity, and temperature.

The amount of astronomical data is massive and has been growing rapidly in recent years due to improved storage capacity and scientific instruments~\cite{zhang2015astronomy}. The rate of growth of this data is outpacing the computational improvements from Moore's Law~\cite{brescia2012extracting}. As a result, better algorithms are required to keep up with this growth of data. This has prompted the emergence of astroinformatics, a new discipline at the intersection of astronomy and computer science~\cite{brescia2012extracting}.

The Sloan Digital Sky Survey (SDSS)~\cite{blanton2017sdss}, the first in a series of survey telescopes, is designed to catalog vast numbers of celestial objects. SDSS comprises several projects, including the Apache Point Observatory Galactic Evolution Experiment (APOGEE)~\cite{majewski2017apogee}, which by 2020 will have cataloged over 263,000 stars in the Milky Way galaxy. SDSS will be succeeded by the Large Synoptic Survey Telescope (LSST)~\cite{ivezic2019lsst} which is expected to catalog 37 billion stars and galaxies over ten years, producing 15 terabytes of data every night. Other large astronomical data sets are becoming available, such as the recent release of the National Optical Astronomy Observatory's All-Sky Survey~\cite{nidever2018first}, which contains 2.9 billion unique objects.

Prior work by Yu, et al. demonstrated that a flat clustering approach could accelerate approximate search on biological and biochemical data sets~\cite{yu2015entropy}. As the growth of biological data has also outpaced Moore's law~\cite{berger2016computational}, we also build on these results by demonstrating the applicability of CHESS to a biological data set. Metagenomics is the study of entire microbial communities in environments such as seawater, soil, or the human gut. The compositional and functional analysis of the human gut microbiome is essential in the study of human health and disease~\cite{gill2006metagenomic, arumugam2011enterotypes, yatsunenko2012human}. One common approach to identifying what species or subspecies of bacteria are present is 16S ribosomal gene amplicon sequencing~\cite{jovel2016characterization}. The ribosome, essential to all known life, comprises both protein and RNA elements; the 16th component of the small ribosomal subunit (16S) is highly conserved among bacteria, but has enough variation to make it a useful ``fingerprint'' to identify bacterial species~\cite{langille2013predictive}. The GreenGenes project~\cite{desantis2006greengenes} provides a multiple-sequence-alignment of over one million bacterial 16S sequences.

Recent approaches to dealing with the exponential growth of data have included locality-sensitive hashing~\cite{indyk1999sublinear}, clever text indices such as FM Index~\cite{simpson2010efficient}, and a recent paper introducing ``entropy-scaling search'' for large biological data sets~\cite{yu2015entropy}. The key contribution of~\cite{yu2015entropy} is an asymptotic complexity for $\rho$-nearest neighbors search. This asymptotic complexity is given by:

\begin{gather}
    O\Bigg(
    \underbrace{k}_{\textrm{metric entropy}} +
    \overbrace{\left|B_D(q,r)\right|}^{\textrm{output size}}
    \underbrace{\left(\frac{r+2r_c}{r}\right)^d}_{\textrm{scaling factor}}
     \Bigg)
     \label{es-eqn}
\end{gather}

where $k$ denotes the metric entropy of the data at cluster radius $r_c$ (equivalently, the number of clusters), $B_D(q,r)$ denotes the number of data points in a ball of radius r around a query $q$ (i.e. the output size of the search), and $d$ denotes the fractal dimension in the range from query radius to cluster radius. Notably, $n$ (the size of the database) does not appear in the complexity (though $B_D(q,r)$ is unlikely to be independent of the size and density of the data). Entropy-scaling search demonstrates particular effectiveness when data appear confined to a lower-dimensional manifold of a high-dimensional space, as hypothesized in~\cite{yu2015entropy} and further explored in~\cite{berger2016computational}.

The tasks of searching, analyzing, and storing the astronomical and biological data discussed above pose a true ``big data'' problem. Approximate search is useful for ``guilt by association'' to inexact matches that have been well studied, and is also a building block for the ubiquitous k-nearest neighbors (KNN) algorithm~\cite{cover1967nearest}. In this paper, we focus on the specific problem of approximate search ($\rho$-nearest neighbors search). Building on the concept of ``entropy-scaling search'' as applied to biological and biochemical data~\cite{yu2015entropy}, we extend entropy-scaling $\rho$-nearest neighbors search to a hierarchical clustering approach, with similarities to binary search, called CHESS (Clustered Hierarchical Entropy-Scaling Search). In doing so, we gain an expected logarithmic improvement in the coarse search, bringing the $k$ term to $\log_2 k$. We demonstrate this approach on the GreenGenes biological data set and the Sloan Digital Sky Survey's APOGEE astronomical data set. It should be noted CHESS is applicable to any problem with a well-defined distance function. In the case of astronomical data, we compare the performance of CHESS to FALCONN~\cite{razenshteyn2015falconn}, a state-of-the-art locality sensitive hashing library. CHESS also allows for lossless compression for storage and transmission of data.

Given the rise of ``big data,'' there has been an interest in sublinear-time algorithms~\cite{indyk1999sublinear,bhattacharya2015space,kane2010exact}. With CHESS, we demonstrate a search algorithm whose asymptotic complexity \textit{is not a function of data set size} but rather geometric properties of the data. CHESS's advantages over na\"ive search will grow as data sets grow in size. We demonstrate that CHESS is an order-of-magnitude (or more) faster than na\"ive search.

\section{Methods}

In this manuscript, we are primarily concerned with the problem of $\rho$-nearest neighbors search in a finite-dimensional (typically high-dimensional) vector space. 

For our purposes, a \textit{data point} is one datum or observation; it might represent the electromagnetic spectrum of a star, the entire genome of an organism, or any other entity on which a distance function is defined.

A \textit{distance function} is any function $d: \mathbb{D} \times \mathbb{D} \to \mathbb{R_+}$, where $\mathbb{D}$ is the set of data points. A distance function may also map to a subset of the nonnegative real numbers (for instance, cosine distance). A distance function that obeys the triangle inequality is also a \textit{distance metric}. Sensible distance functions are chosen for the data: for spectral data, both Euclidean ($L2$) and cosine distance are used; for biological sequence data, Levenshtein edit distance or Hamming distance make sense~\cite{needleman1970general}.

We distinguish the \textit{embedding dimension} from another concept, the discrete \textit{fractal dimension} of the data at some particular length scale. We define \textit{local fractal dimension} as:

\begin{gather}
    \log_2\bigg(\frac{|B_D(q, r_1)|}{|B_D(q, r_2)|}\bigg)
    \label{fractal-dimension}
\end{gather}

where $B_D(q,r)$ is the set of points contained in a ball on the dataset $D$ of radius $r$ centered on a point $q$; here, fractal dimension is computed for a radius $r_1$ and a smaller radius $r_2=\frac{r_1}{2}$

For instance, the SDSS's APOGEE data are embedded in $\mathbb{R}_+^{8575}$ (they are nonnegative real-valued vectors of length 8,575) and they do not uniformly occupy that space. This is due to physical constraints (namely, the laws that govern stellar fusion and spectral emission lines); as famouly illustrated by the Hertzsprung-Russel diagram~\cite{rosenberg1910zusammenhang} plotting luminosity against temperature. The notion of high-dimensional data constrained to a lower-dimensional manifold is known as the \textit{manifold hypothesis}~\cite{fefferman2016testing}.

We define a \textit{cluster} as a set of points with a \textit{center}, a \textit{radius}, and an approximation of the \textit{local fractal dimension}. The \textit{center} is the geometric median of the set of points in the cluster, and so is a real data point. The \textit{radius} is the maximum distance from the points to the center. The \textit{local fractal dimension} is estimated at the cluster radius and half the cluster radius. Each \textit{cluster} (unless it is a \textit{leaf cluster}) has two \textit{child clusters} in much the same way as a node in a binary tree has two child nodes.

We define the \textit{metric entropy} of a data set under a hierarchical clustering scheme as a refinement of~\cite{yu2015entropy}, where metric entropy for a given cluster radius $r_c$ was the number of clusters of that radius needed to cover all data. Here, we use a binary, divisive clustering approach, but with early-stopping criteria; clusters smaller than a certain threshold are not further split. Since we frame the asymptotic complexity of search in terms of the number of leaf clusters, the metric entropy is best thought of in terms of the number of leaf clusters, which is an emergent property for any given data set. Divisive clustering is chosen because we want a hierarchical ``walk'' of the data during search; determining an optimal clustering is not the goal.

\subsection{Data Sets}

\subsubsection{Sloan Digital Sky Survey}

The Sloan Digital Sky Survey~\cite{blanton2017sdss} Apache Point Observatory Galaxy Evolution Experiment (APOGEE)~\cite{majewski2017apogee} data set contains near-infrared spectra of approximately 130,000 stellar objects in the Milky Way galaxy. Each spectral datum describes a star's flux in Janskys ($1~Jy = 10^{-26}~W\cdot m^{-2}\cdot Hz^{-1}$) at each of $8,575$ wavelengths ranging from $\lambda = 1.51~\mu m$ to $1.70~\mu m$. Thus, each datum is a real-valued vector in $\mathbb{R}_+^{8575}$. These near-infrared spectra are used by astronomers to achieve a better understanding of how our galaxy has evolved. The APOGEE dataset used was downloaded in June 2017 from SDSS Data Release 12~\cite{alam2015eleventh}.

\begin{figure}[!t]
\centering
\includegraphics[width=2.5in]{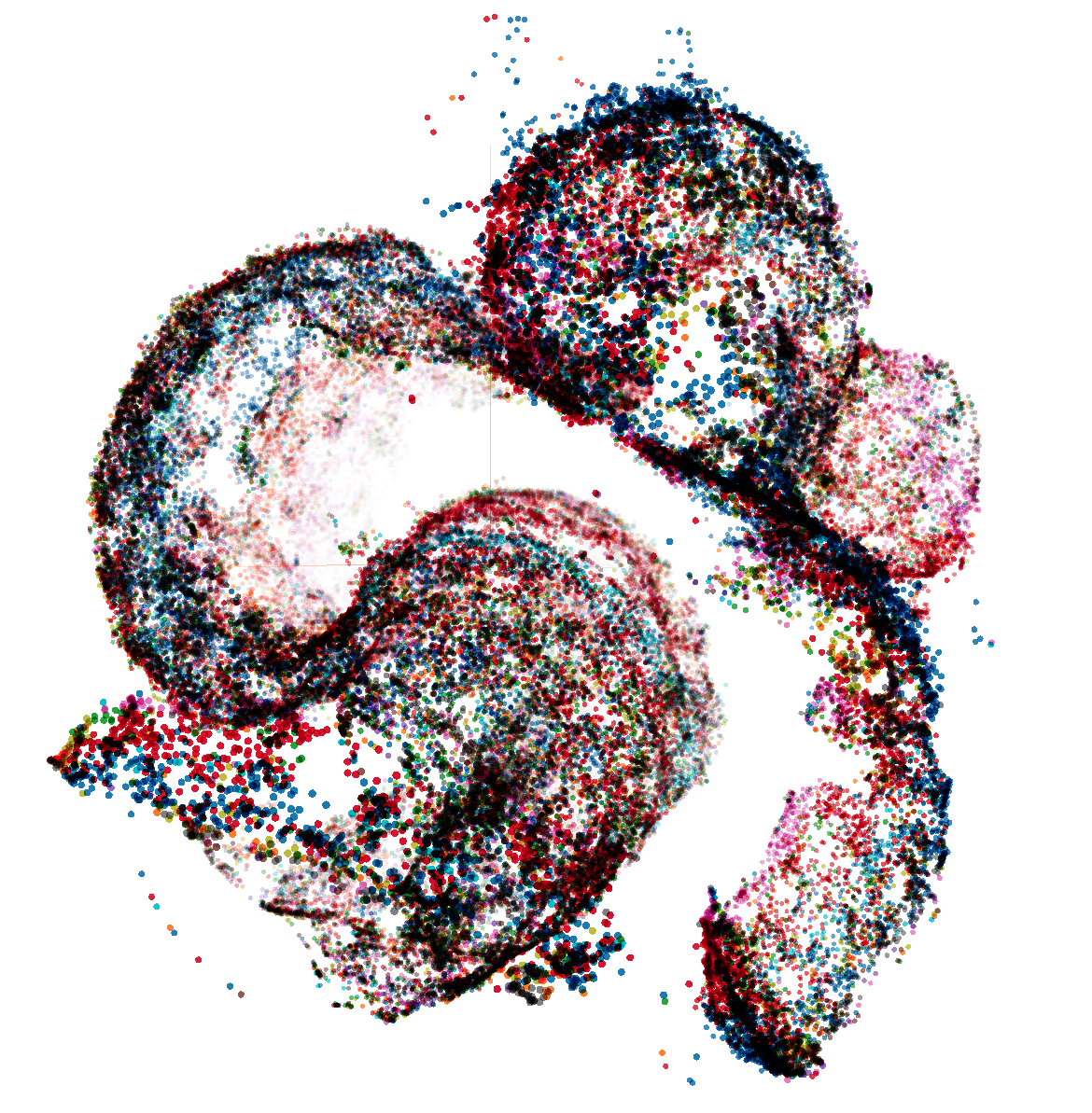}
\caption{T-SNE visualization of the SDSS Apogee data set, using L2 norm as the distance function.}
\label{apogee-l2}
\end{figure}

Figure~\ref{apogee-l2} shows a t-Stochastic Neighbor Embedding (t-SNE)~\cite{maaten2008visualizing} visualization of the APOGEE data, where colors represent leaf clusters of spectra (some colors are reused due to a limited palette). This visualization suggests support for the ``manifold hypothesis.'' Note that while the variation among spectra appears relatively continuous (there are few distinct, isolated clusters), there are large regions of unoccupied space in the reduced-dimension representation. In contrast, uniformly distributed data, when visualized via t-SNE, appear as a uniform cloud. This apparent constraint to a ``sheetlike'' manifold embedded in the high-dimensional vector space suggests that the APOGEE data is a reasonable candidate for CHESS.

\begin{figure}[!t]
\centering
\includegraphics[width=3in]{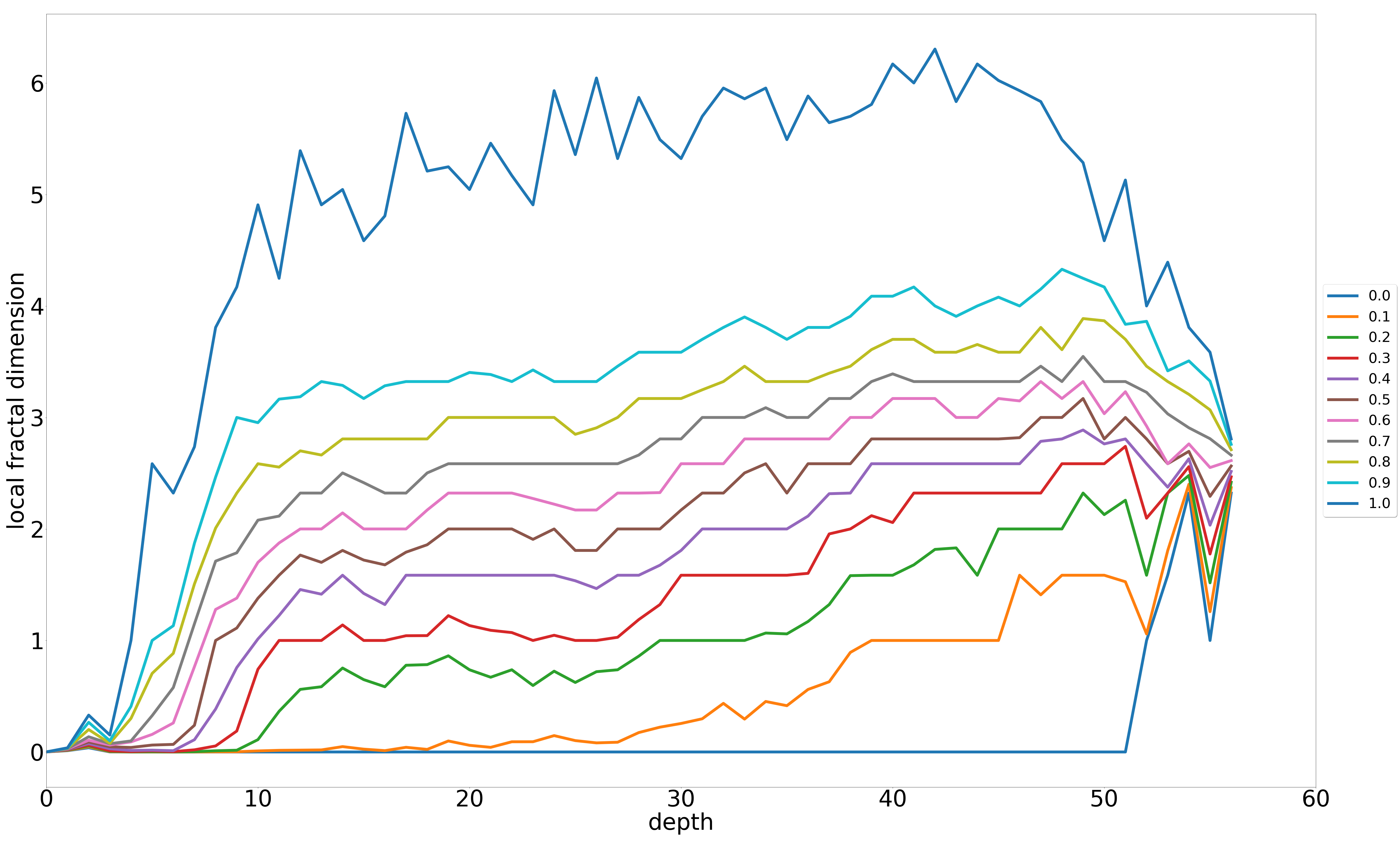}
\caption{Mean fractal dimension of APOGEE clusters as a function of depth when clustered based on L2 norm. Each plotline represents a distinct decile of fractal dimension. Beyond a depth of 56, no clusters are further divided due to the minimum-cluster-cardinality stopping criteria.}
\label{apogee-lfd-l2}
\end{figure}

\begin{figure}[!t]
\centering
\includegraphics[width=3in]{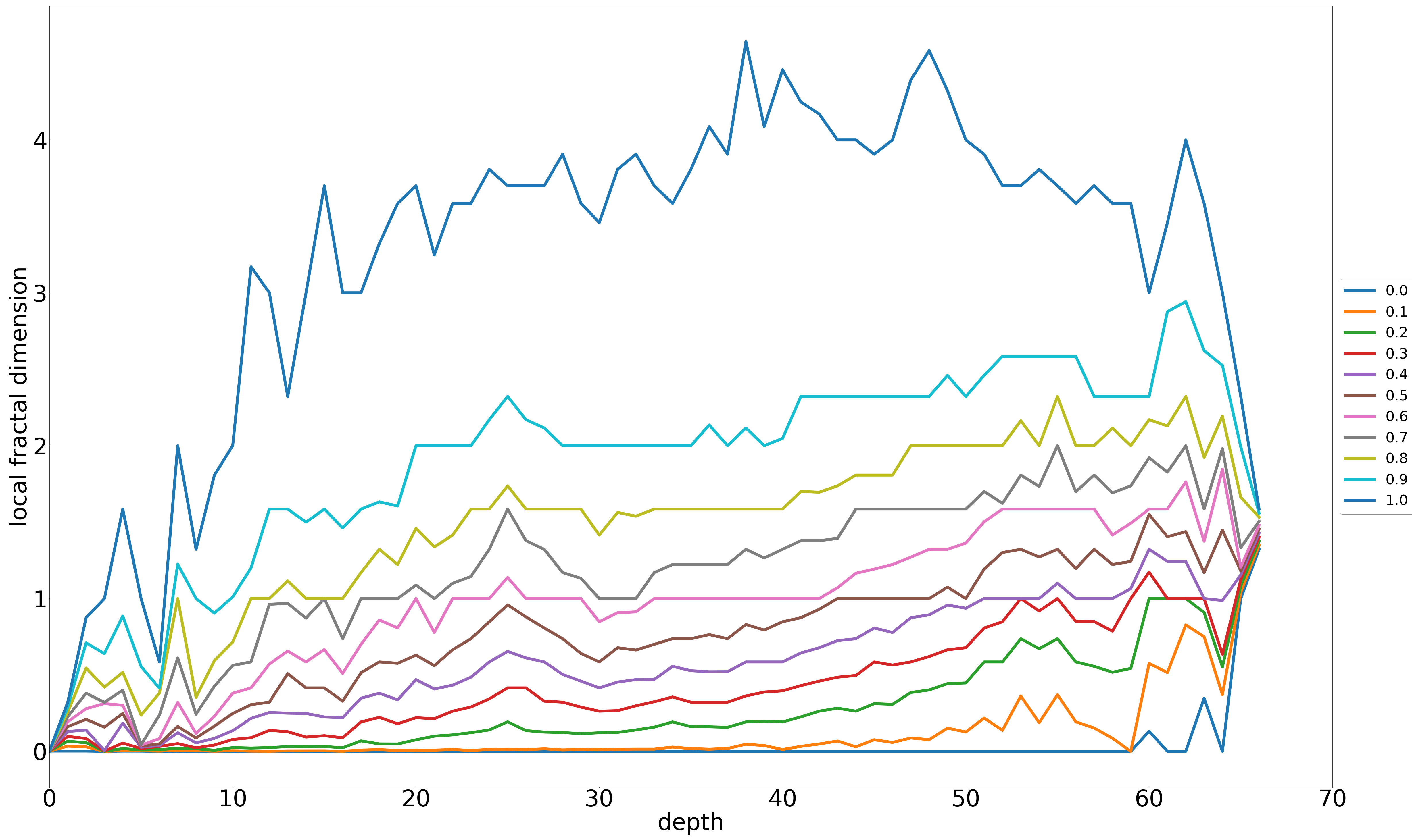}
\caption{Mean fractal dimension of APOGEE clusters as a function of depth when clustered based on cosine distance. Each plotline represents a distinct decile of fractal dimension. Beyond a depth of 66, no clusters are further divided due to the minimum-cluster-cardinality stopping criteria.}
\label{apogee-lfd-cos}
\end{figure}

Figure~\ref{apogee-lfd-l2} shows the mean local fractal dimension of the APOGEE data at each level of hierarchical clustering, under the L2 (Euclidean) distance metric. Each plot line represents a decile of fractal dimension; note that, other than the most extreme 10\% of clusters, virtually all clusters have a local fractal dimension of less than 2. Figure~\ref{apogee-lfd-cos} provides the same visualization under cosine distance and, once again, indicates that most clusters have a local fractal dimension of less than 2.

\subsubsection{GreenGenes bacterial genomes}

The GreenGenes project~\cite{desantis2006greengenes} provides a set of 1,027,383 bacterial 16S sequences. These sequences contain several duplicates so 805,434 unique sequence user used in clustering. Given a metagenomic sample, a bioinformatician might search those million-plus sequences for close matches to the 16S sequences identified in a sample of interest. A typical distance function here would be edit (Levenshtein) distance or Hamming distance, which roughly models the insertions, deletions, and substitutions that happen during the course of evolution~\cite{needleman1970general}. These sequences are already aligned and, thus, form sequences of $\{A,C,G,T,-\}$ (where $-$ represents a gap corresponding to an insertion or deletion during the course of evolution). Thanks to the alignment, every sequence in the set is 7,682 characters long, so Hamming distance is an appropriate metric.

\begin{figure}[!t]
\centering
\includegraphics[width=2.5in]{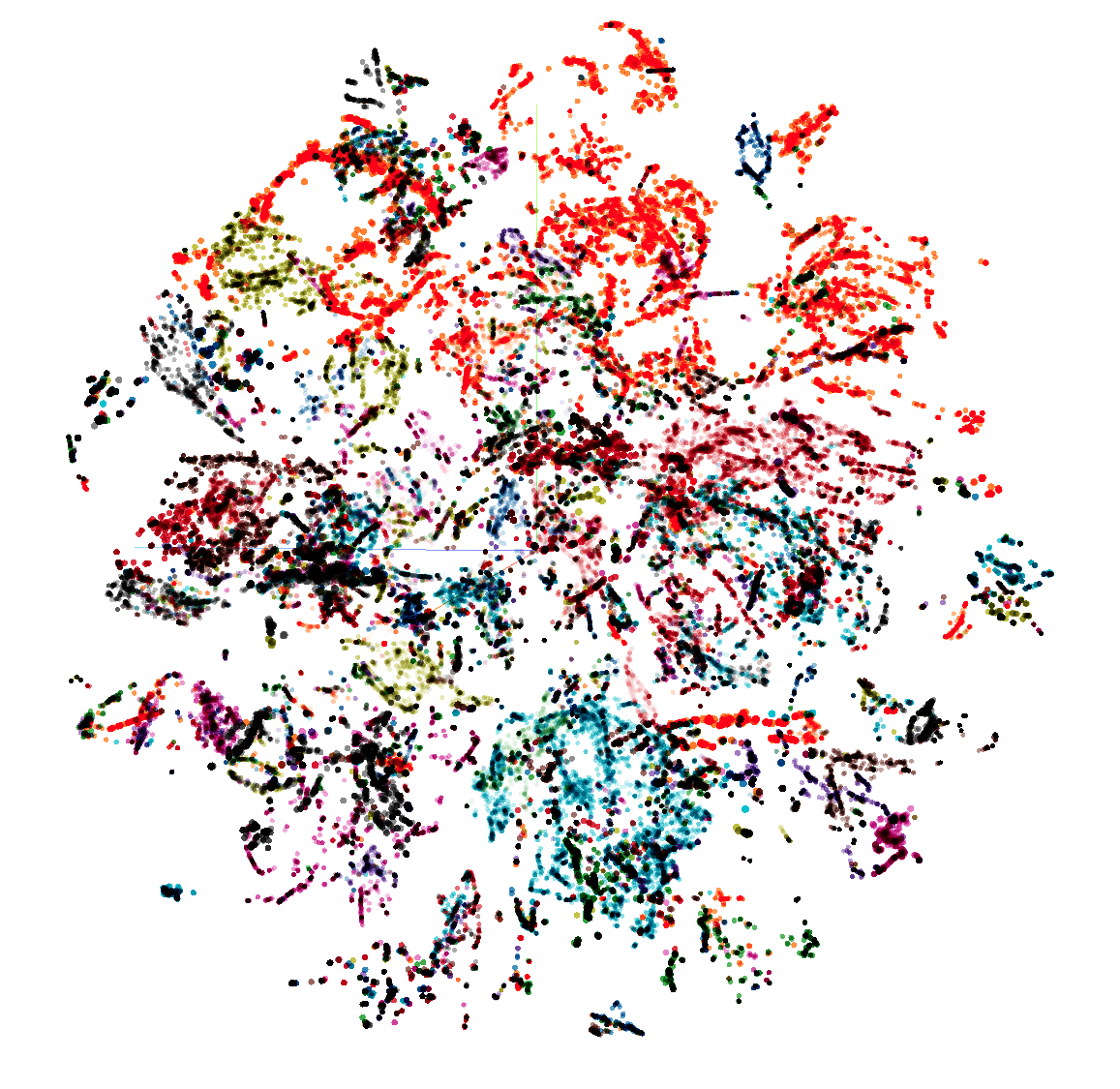}
\caption{T-SNE visualization of the GreenGenes 16S RNA data set, using Hamming distance.}
\label{greengenes-hamming}
\end{figure}

During the course of evolution, organisms adapt to their environment by random mutation and the retention of (or selection for) those mutations that are beneficial. In~\cite{yu2015entropy}, it was hypothesized that this process could result in the low fractal dimension and low metric entropy for which entropy-scaling search is suited. Figure~\ref{greengenes-hamming} shows a t-SNE visualization of the GreenGenes dataset under Hamming distance, where colors indicate the genus of a bacterium (some colors are reused due to a limited palette). Clearly, these data appear fundamentally different from the APOGEE data. There are distinct clusters of species separate from one another, which tend to be grouped by genus, and the continuity seen in the APOGEE data is not present. Like the APOGEE data, entire regions of the embedding space are unoccupied, suggesting that CHESS is well-suited to the GreenGenes dataset.

\begin{figure}[!t]
\centering
\includegraphics[width=3in]{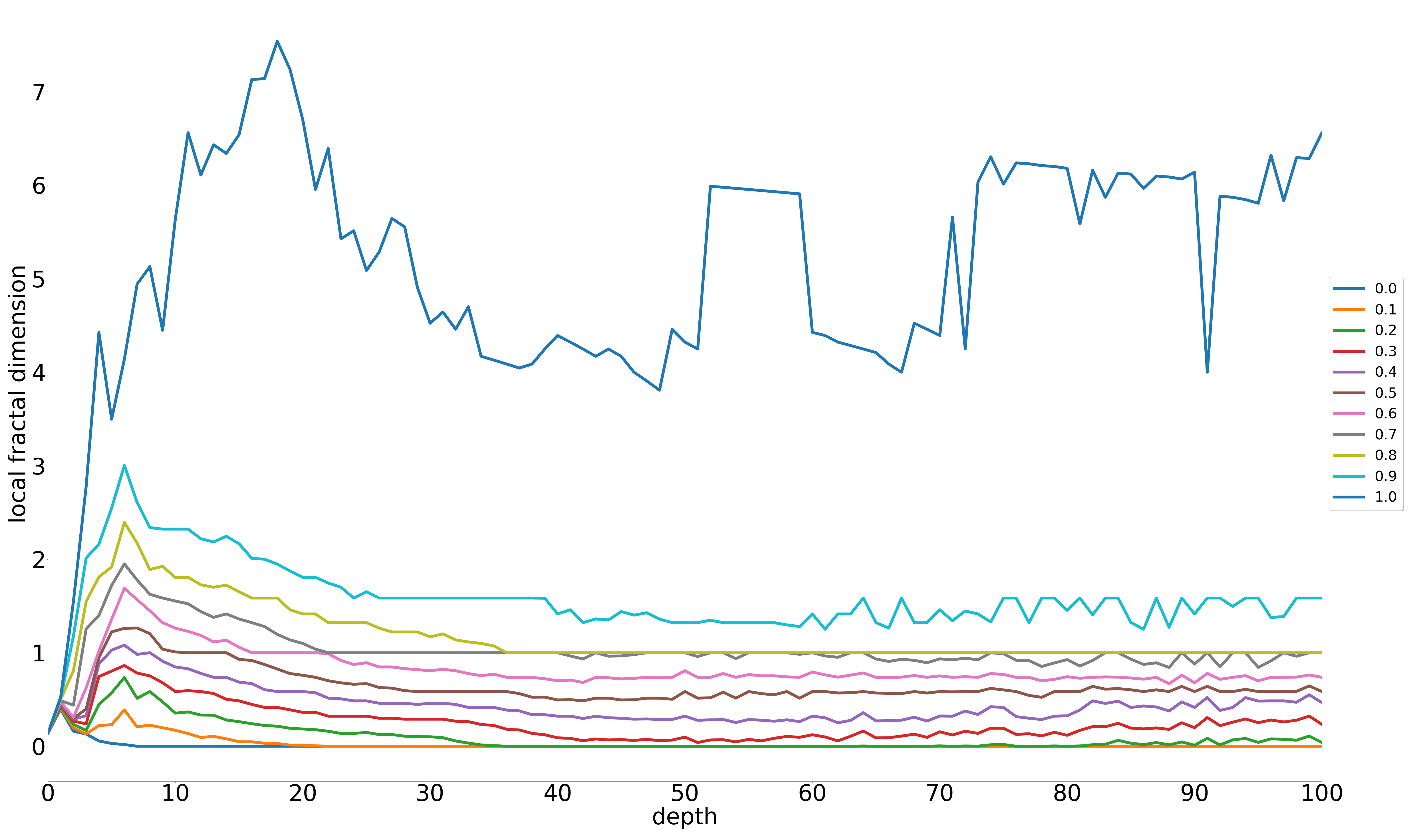}
\caption{Mean fractal dimension of GreenGenes clusters as a function of depth when clustered based on Hamming distance. Each plotline represents a distinct decile of fractal dimension.}
\label{gg-lfd-hamming}
\end{figure}

Figure~\ref{gg-lfd-hamming} shows the local fractal dimension of the GreenGenes data at each level of hierarchical clustering under Hamming distance. Each plot line represents a decile of fractal dimension; once again, other than the most extreme 10\% of clusters, virtually all clusters have a local fractal dimension of less than 2.

For this study, we used the October 2012 release of Greengenes, downloaded in April 2019.

\subsection{Hierarchical clustering and search}

\subsubsection{Clustering}

Hierarchical clustering lends itself to a binary search on clusters, followed by one or more ``fine'' (linear) searches inside clusters. As such, the $k$ term in the asymptotic complexity from flat-clustered search becomes $\log_2 k$. However, the fine search term is more difficult to quantify, as cluster radius is no longer an input parameter, but instead an emergent property. Unlike classical binary search, there are cases (such as the ``ball'' around a query crossing into multiple child clusters) where multiple branches of the binary tree must be searched. However, under an assumption of low fractal dimension and low metric entropy, where much of the high-dimensional space is unoccupied by data, traversing multiple branches of the tree should be rare. This is demonstrated in Section~\ref{results}.

Hierarchical clustering follows the procedure outlined in Algorithm~\ref{clustering}.

\begin{algorithm} 
\caption{Cluster(data)} 
\label{clustering} 
\begin{algorithmic} 
    
    \REQUIRE $maxdepth > 0$
    \REQUIRE $minsize > 0$
    \STATE $d \Leftarrow 0$
    \STATE $clusters = \{\}$
    \STATE $n = |data|$
    \WHILE{$d \le maxdepth$}
        \STATE $numseeds \Leftarrow \sqrt{n}$
        \STATE $seeds \Leftarrow numseeds$ chosen randomly from $data$
        \STATE $l, r \Leftarrow \{l, r | l,r \in seeds \land l,r = \argmax d(x,y) | x,y \in seeds\}$
        \STATE $clusters[l] \Leftarrow \{x | x \in data \land d(l,x) \le d(r,x)\}$
        \STATE $clusters[r] \Leftarrow \{x | x \in data \land d(r,x) < d(l,x)\}$
        \IF{$|clusters[l]| > minsize$}
            \STATE Cluster(clusters[l])
        \ENDIF
        \IF{$|clusters[r]| > minsize$}
            \STATE Cluster(clusters[r])
        \ENDIF
    \ENDWHILE
\end{algorithmic}
\end{algorithm}

Hierarchical clustering is essentially a recursive procedure that relies on some distance function $d(x,y)$. In order to cluster $n$ data points, $\sqrt{n}$ are chosen at random as possible cluster centers (seeds). All pairwise distances are computed among these $\sqrt{n}$ seeds, and the two furthest seeds are chosen as the ``left'' and ``right'' cluster centers. All $n$ data points are then partitioned into ``left'' and ``right'' ``child'' clusters by whichever seed they are closer to. This procedure recurses until a user-specified maximum depth is reached. Any cluster containing fewer than a user-specified minimum count (typically 10) is not recursively clustered further.

\subsubsection{Search}

Rather than an exact search (which may not return any hits), we conduct an approximate ($\rho$-nearest neighbor) search. Given a hierarchical clustering, the search procedure begins much like classical binary search.  During this search, it is possible that both children of a cluster contain valid hits. Thus, a search radius of $r + r_{c_i}$ (the query radius plus the cluster radius for a cluster $c_i$) is used to determine whether a subtree might contain valid hits. Search proceeds recursively down the binary tree of clusters, exploring any child of a cluster if that child's cluster center is within $r+r_c$ of the query, where $r_c$ is the radius of that child. Once leaves are reached, the union of all leaves identified as containing candidate hits are searched exhaustively. The procedure is outlined in Algorithm~\ref{search}.

\begin{algorithm} 
\caption{Search(q,r,clusters)} 
\label{search} 
\begin{algorithmic} 
    
    \REQUIRE $r \ge 0$
    \REQUIRE $clusters \ne \emptyset$
    \STATE $results \Leftarrow \{\}$
    \IF{$clusters.left$}
        \IF{$d(q, clusters.l.center) \le r + clusters.l.radius$}
            \STATE Search(q,r,clusters.l)
        \ENDIF
    \ENDIF
    \IF{$clusters.right$}
        \IF{$d(q, clusters.rt.center) \le r + clusters.rt.radius$}
            \STATE Search(q,r,clusters.rt)
        \ENDIF
    \ENDIF
    \IF{$\lnot clusters.l \land \lnot clusters.rt$}
        \FOR{$p \in clusters.points$}
            \IF{$d(q,p) \le r$}
                \STATE $results \Leftarrow r$
            \ENDIF
        \ENDFOR
    \ENDIF
    \RETURN $results$        

\end{algorithmic}
\end{algorithm}

\subsection{Data compression}\label{compression}

Each leaf cluster contains a cluster center (a real data point, rather than a centroid). Therefore, each member of that cluster can be represented as the set of differences from that center. The optimal representation of these differences depends on the nature of the data as well as the distance function. In the case of string data (such as biological sequence data) under a Hamming distance metric, the representation is naturally a set of edits to transform the cluster center into any data point in the cluster. \textit{The number of these edits is bounded by the cluster radius}. In the case of astronomical spectra, the data are real-valued vectors in ($\mathbb{R}_+^{8575}$) and so a compact encoding appears elusive. This is because the upper bound difference of one cluster radius from the center could be distributed in any manner among the 8,575 bins.

However, the underlying APOGEE data is not of arbitrary precision. Due to the physical limitations inherent in measurement, these values are quantized~\cite{goldberg1986image} and this makes an integer encoding of differences possible. Consider, by way of analogy, a meter rule that is marked in increments of $1 mm$. Any length measured by the meter rule is measured as some integer multiple of $1 mm$ even though the actual length is some real number. Any difference of less than $1 mm$ between two lengths cannot be measured and is ignored. Lengths measured by such a meter rule can, thus, be thought to be quantized by $1 mm$ sized quanta. The telescope and spectrometer used for APOGEE have an analogous limitation. APOGEE reports that it can detect stars of magnitude $H = 12.2$~\cite{majewski2017apogee} (where $H = -2.5 \log_{10}(I)$ and $I$ is the intensity of the star). This leads to quanta of size $10^{-12.2/2.5}$.

For lossless compression of clustered data, we calculate the difference between each point in a cluster and the cluster center. This difference is then quantized to give an integer encoding. These encodings can then be compressed using any suitable compression library. We used the \texttt{zipfile} library in Python for our implementation.

\subsection{Benchmarks}

\subsubsection{Sloan Digital Sky Survey APOGEE}

Fifty randomly-selected data points were held out from the data prior to hierarchical clustering. Each of those points was treated as a query, and the data set was searched to return all spectra within a specific radius of the query. For L2 (Euclidean) distance, these radii were 2,000 and 4,000. We measured the wall-clock time for each query, as well as the number of distance comparisons required to perform the search. We report the mean and standard deviation for distance comparisons, search time (in seconds), the fraction of the dataset that was searched, and the mean speedup factor compared to a na\"ive, linear search implementation. This linear search must make $n$ comparisons given a database of size $n$.

All benchmarks were conducted on an Intel Xeon E5-2690 v4 2.60GHz (single threaded), with 512GB RAM and CentOS 7 Linux with kernel 3.10.0-862.9.1.el7.x86\_64. FALCONN is not versioned, but commit \texttt{a4c0288edb2575b0306c5f6b4ab1f559b1e2a} from GitHub was used.

FALCONN is written in C++ (and heavily optimized for CPUs) while CHESS is written in Python. As such, wall-clock time comparisons are not truly fair; a GPU-optimized or multithreaded FALCONN would likely provide improved performance, while CHESS rewritten in C++ would likely boast improved all-round performance. To remedy this inconsistency, we report the \textit{number of distance comparisons made} by the two methods while conducting the same search on the same dataset. We used Valgrind's \texttt{callgrind} tool~\cite{nethercote2007valgrind} to produce a performance profile of FALCONN's underlying C++ library, and we used internal source-code instrumentation to profile CHESS.

\subsubsection{GreenGenes}

Fifty randomly-selected data points were held out from the data prior to hierarchical clustering. Each of those points was treated as a query, and the data set was searched to return all 16S sequences within a specific radius of the query. We used Hamming distances corresponding to 99.9\% and 99\% sequence identity (note that at 95\% sequence identity a query returns, on average, nearly 75\% of the data set, while at 90\% sequence identity a query would return almost the entire dataset, so we expect no improvement over na\"ive search). We measured the wall-clock time for each query, as well as the number of distance comparisons required to perform the search. We report the mean and standard deviation for comparisons, search time (in seconds), the fraction of the dataset that was searched, and the mean speedup factor compared to a na\"ive, linear search implementation.

The GreenGenes benchmarks for CHESS were conducted on a Xeon E5-2620 v4 2.10GHz with 64GB RAM, hosted on Ubuntu Linux with kernel 4.15.0-54-generic, Python 3.6.8.

FALCONN only supports L2 and cosine distances, not Hamming distance, so it is not directly comparable on the GreenGenes data set. Thus, we did not benchmark FALCONN on GreenGenes.

\section{Results}\label{results}

\subsection{Asymptotic complexity}

To analyze the computational complexity of hierarchical clustering, consider the top level of the hierarchy. There are $n$ data points, from which $\sqrt{n}$ seeds are chosen. All pairwise distances between these seeds are computed, i.e. $n$ comparisons are made. The furthest pair of points are chosen as the centers of child-clusters. Every data point except the two chosen centers is then compared to both centers; i.e., an additional $2n$ comparisons. In general, at depth $d$ of the hierarchy, we have $\frac{n}{2^d}$ points per cluster (assuming a perfectly balanced clustering, which will turn out to not be required), $\sqrt{\frac{n}{2^d}}$ seeds, $\frac{n}{2^d}$ pairwise seed comparisons, and $2 \times \frac{n}{2^d}$ additional comparisons when each other data point is compared to the two cluster centers, for a total of $3 \times \frac{n}{2^d}$ distance comparisons. At each level, there are $2^d$ clusters, so the total number of distance comparisons per level is $3n$. Thus, the asymptotic complexity in terms of distance comparisons for a cluster tree of depth $d$ is $O(dn)$.

Since hierarchical entropy-scaling search begins as a search over a binary tree, it would appear that the coarse search (the tree traversal) would take $O(\log_2 k)$ time, if there are $k$ leaf clusters. However, there is a complication: since this is approximate rather than exact search, a nonzero query radius, $r$, is used. The ``ball'' of possible points around the query may overlap multiple branches of the tree. While it is clear that some of the search space is pruned during the tree traversal, there is no guarantee that the fine search will be over only one leaf cluster.

Consider a clustering using L2 norm. Suppose that the data comprise a one-dimensional manifold in a two-dimensional embedding. Then, every level of the binary tree represents some line dividing the parent cluster into two (not necessarily equally-sized) clusters. If a parent cluster is not identified as a target for search, neither of its children will be. However, it is possible that \textit{both} children of a given parent cluster could be considered. Once cluster radius has decreased to be smaller than the search radius, there is no advantage to further tree traversal, and the entire subtree must be searched. However, above the level of the tree where search radius approximates cluster radius, at most both children of an internal cluster node need to be searched; even in this instance, only half of their descendents would need to be searched, because of the two-dimensional nature of the data set. Thus, the coarse search (in one dimension) requires only a constant factor of 2; $O(2 \log_2 k) = O(\log_2 k)$. However, as dimensionality increases the argument requires using the geometry of an $n-$sphere.

Recall that our clustering approach chooses (or estimates) the two furthest points in a cluster to determine the cluster centers at the next level of the hierarchy. First, consider data in one dimension. The very first cut into two clusters results in a boundary on which a query with nonzero radius might hit results from both clusters. However, any further division of those clusters separates the data into non-adjacent regions, so continued binary clustering and then search will result in a pruning of the search space. Next, consider data in two dimensions. Without loss of generality, consider those data uniformly occupying the interior of a circle around some origin. A first cut will separate the data into two approximate semicircles. But, the furthest points on each semicircle will likely be 180 degrees apart, along the diameter of the cut. Thus, a second cut will separate each semicircle into a quarter-circle, which contains a right triangle. Thus, the furthest points in each quarter-circle could still be at the two intersections of the triangle with the circumference of the circle. It is only after this point that, under a uniform distribution of data, the third cut will separate each quarter-circle into an eighth-circle, leading the fourth cut to be orthogonal to a radius, separating ``outer'' points from ``inner'' points and guaranteeing that no query with radius smaller than the cluster radius would require searching inside every cluster.

Taken to three dimensions, orthogonal cuts are only guaranteed once the wedges of the sphere are less than 60 degrees in both $\phi$ and $\theta$, which requires six cuts. Generalizing to the $n-$sphere, described in polar coordinates as containing angles $\phi_1 \cdots \phi_n$, the first $3\times(d-1)$ cuts are not guaranteed to be orthogonal to a radius. If our data were uniformly distributed in a high-dimensional space, this would bode poorly for CHESS. 

Of course, with the APOGEE data set, we have not two dimensions but 8,575. By the curse of dimensionality, we might need to explore many thousands of ``neighboring'' tree branches. How can this explosion be reconciled with the empirically faster search as tree depth increases? There is no such explosion because the number of neighboring branches that may need to be searched grows not with the \textit{embedding} dimension (i.e. the dimensionality of each data point, 8,575 for APOGEE and 7,682 for GreenGenes) but with the \textit{local fractal dimension} on the length scale of the cluster and search radii. As in \cite{yu2015entropy}, we use Equation~\ref{fractal-dimension} to estimate the local fractal dimension between two radii at a given point. As illustrated in Figures~\ref{apogee-lfd-l2},~\ref{apogee-lfd-cos},~and~\ref{gg-lfd-hamming}, this fractal dimension is low, typically less than 2, except for the most extreme decile of clusters. Thus, \textit{empirically} for both the APOGEE and GreenGenes data sets, the multiplier is bounded by 4, which means that once clustering exceeds a depth of 3, we can claim an effective asymptotic complexity of:

\begin{gather}
    O\Bigg(
    \underbrace{\log_2 k}_{\textrm{metric entropy}} +
    \overbrace{\left|B_D(q,r)\right|}^{\textrm{output size}}
    \underbrace{\left(\frac{r+2\hat{r_c}}{r}\right)^d}_{\textrm{scaling factor}}
     \Bigg)
     \label{hierarchical-complexity}
\end{gather}

where $\hat{r_c}$ is the \textit{mean} cluster radius of leaf clusters.

Similar to the original entropy-scaling search~\cite{yu2015entropy}, this approach is ill-suited to data that are not constrained to low-dimensional manifolds in their higher-dimensional embedding. On the other hand, when data exhibit such constraints, entropy-scaling search is appropriate.

\subsection{Benchmark Results}

\begin{figure}[!t]
\centering
\includegraphics[width=3in]{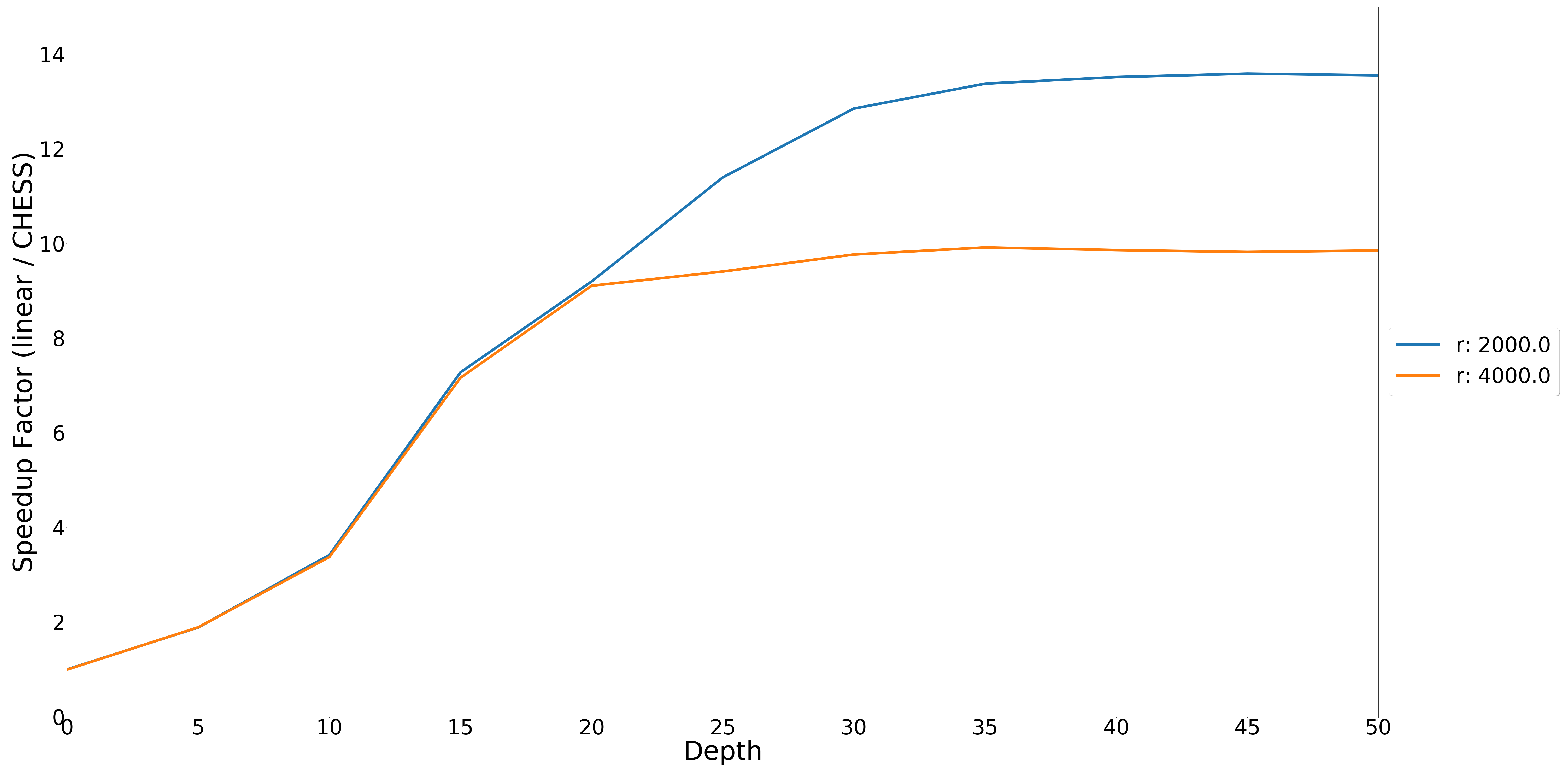}
\caption{Speedup Factor of CHESS on APOGEE vs. Na\"ive Search Under L2 Norm as a Function of Search Depth}
\label{speedup_factor_l2_png}
\end{figure}

On the SDSS's APOGEE data set, using L2 norm, there is a clear reduction in search time as the maximum tree depth increases. Table~\ref{l2} shows the number of comparisons, search time in seconds, what fraction of the data needed to be searched, and the speedup factor vs. na\"ive linear search for L2 norm. Figure~\ref{speedup_factor_l2_png} illustrates the speedup as a function of depth.

\begin{table}[!t]
\renewcommand{\arraystretch}{1.3}
\caption{CHESS performance on APOGEE vs. na\"ive, L2 norm}
\label{l2}
\centering
\begin{tabular}{|c|c|c|c|c|c|c|c|}
\hline
\bfseries Depth & \multicolumn{2}{c|}{\textbf{Comparisons}} & \multicolumn{2}{c|}{\textbf{Search}} & \multicolumn{2}{c|}{\textbf{Fraction}} & \bfseries Speedup\\
 & \multicolumn{2}{c|}{$\times 10^{4}$} & \multicolumn{2}{c|}{\textbf{Time (s)}} & \multicolumn{2}{c|}{\textbf{Searched}} & \bfseries Factor\\
\hline
 & \bfseries $\mu$ & \bfseries $\sigma$ & \bfseries $\mu$ & \bfseries $\sigma$ & \bfseries $\mu$ & \bfseries $\sigma$ & \bfseries $\mu$\\
\hline
\multicolumn{3}{|l}{\textbf{r = 2,000}} & \multicolumn{5}{|l|}{\textbf{Output Size: $\mu=482~ \sigma=1,004$}}\\
\hline
10 & 7.79 & 2.86 & 5.03 & 1.82 & 0.63 & 0.23 & 3.42\\
20 & 3.30 & 1.88 & 2.38 & 1.22 & 0.27 & 0.15 & 9.20\\
30 & 1.44 & 1.20 & 1.57 & 0.94 & 0.12 & 0.10 & 12.85\\
40 & 0.73 & 0.69 & 1.40 & 0.88 & 0.06 & 0.06 & 13.52\\
50 & 0.68 & 0.63 & 1.38 & 0.87 & 0.05 & 0.05 & 13.55\\

\hline
\multicolumn{3}{|l}{\textbf{r = 4,000}} & \multicolumn{5}{|l|}{\textbf{Output Size: $\mu=4,563~ \sigma=5,803$}}\\
\hline
10 & 7.87 & 2.86 & 5.04 & 1.81 & 0.64 & 0.23 & 3.37\\
20 & 3.55 & 1.99 & 2.57 & 1.30 & 0.29 & 0.16 & 9.11\\
30 & 1.97 & 1.47 & 1.96 & 1.04 & 0.16 & 0.12 & 9.77\\
40 & 1.54 & 1.22 & 1.84 & 0.99 & 0.12 & 0.10 & 9.86\\
50 & 1.54 & 1.21 & 1.84 & 0.98 & 0.12 & 0.10 & 9.85\\
\hline
\end{tabular}
\end{table}

On the APOGEE data set using cosine distance, there is no clear reduction in search time until CHESS begins to significantly prune the search space, which requires a much higher depth than the L2 norm. Table~\ref{cos} shows benchmark results under cosine distance at depths of 10 through 50 with increments of 10. At a depth of 30, CHESS begins to show moderate speedup, leveling off by depth 45. The greater depth requirement appears to be a result of the fact that cosine distance cannot make \textit{orthogonal} cuts to the space the way L2 norm or Hamming distance can. While on a circle, with angle $\theta$, a query with small radius can hit at most two clusters. On a 3-sphere, with angles $\phi$ and $\theta$, a query could hit $2^{d-1}=4$ clusters. Thus, it takes greater cluster depth to begin to see significant pruning of the search space.

\begin{table}[!t]
\renewcommand{\arraystretch}{1.3}
\caption{CHESS performance on APOGEE vs. na\"ive, Cosine Distance}
\label{cos}
\centering
\begin{tabular}{|c|c|c|c|c|c|c|c|}
\hline
\bfseries Depth & \multicolumn{2}{c|}{\textbf{Comparisons}} & \multicolumn{2}{c|}{\textbf{Search}} & \multicolumn{2}{c|}{\textbf{Fraction}} & \bfseries Speedup\\
 & \multicolumn{2}{c|}{$\times 10^{4}$} & \multicolumn{2}{c|}{\textbf{Time (s)}} & \multicolumn{2}{c|}{\textbf{Searched}} & \bfseries Factor\\
\hline
 & \bfseries $\mu$ & \bfseries $\sigma$ & \bfseries $\mu$ & \bfseries $\sigma$ & \bfseries $\mu$ & \bfseries $\sigma$ & \bfseries $\mu$\\
\hline
\multicolumn{3}{|l}{\textbf{r = 0.0005}} & \multicolumn{5}{|l|}{\textbf{Output Size: $\mu=3~ \sigma=8$}}\\
\hline
10 & 12.24 & 0.02 & 7.75 & 0.09 & 0.99 & 0.00 & 1.00\\
20 & 10.82 & 1.08 & 7.58 & 0.73 & 0.88 & 0.09 & 1.04\\
30 & 4.19 & 2.11 & 4.93 & 1.97 & 0.34 & 0.17 & 2.06\\
40 & 0.88 & 0.72 & 3.85 & 1.92 & 0.07 & 0.06 & 2.76\\
50 & 0.62 & 0.49 & 3.83 & 1.93 & 0.05 & 0.04 & 2.78\\
\hline
\multicolumn{3}{|l}{\textbf{r = 0.001}} & \multicolumn{5}{|l|}{\textbf{Output Size: $\mu=73~ \sigma=163$}}\\
\hline
10 & 12.24 & 0.02 & 7.78 & 0.09 & 0.99 & 0.00 & 0.99\\
20 & 10.85 & 1.08 & 7.63 & 0.76 & 0.88 & 0.09 & 1.03\\
30 & 4.40 & 2.21 & 5.19 & 2.14 & 0.36 & 0.18 & 2.03\\
40 & 1.07 & 0.90 & 4.30 & 2.34 & 0.09 & 0.07 & 2.64\\
50 & 0.78 & 0.66 & 4.34 & 2.43 & 0.06 & 0.05 & 2.64\\
\hline
\end{tabular}
\end{table}

\begin{figure}[!t]
\centering
\includegraphics[width=3in]{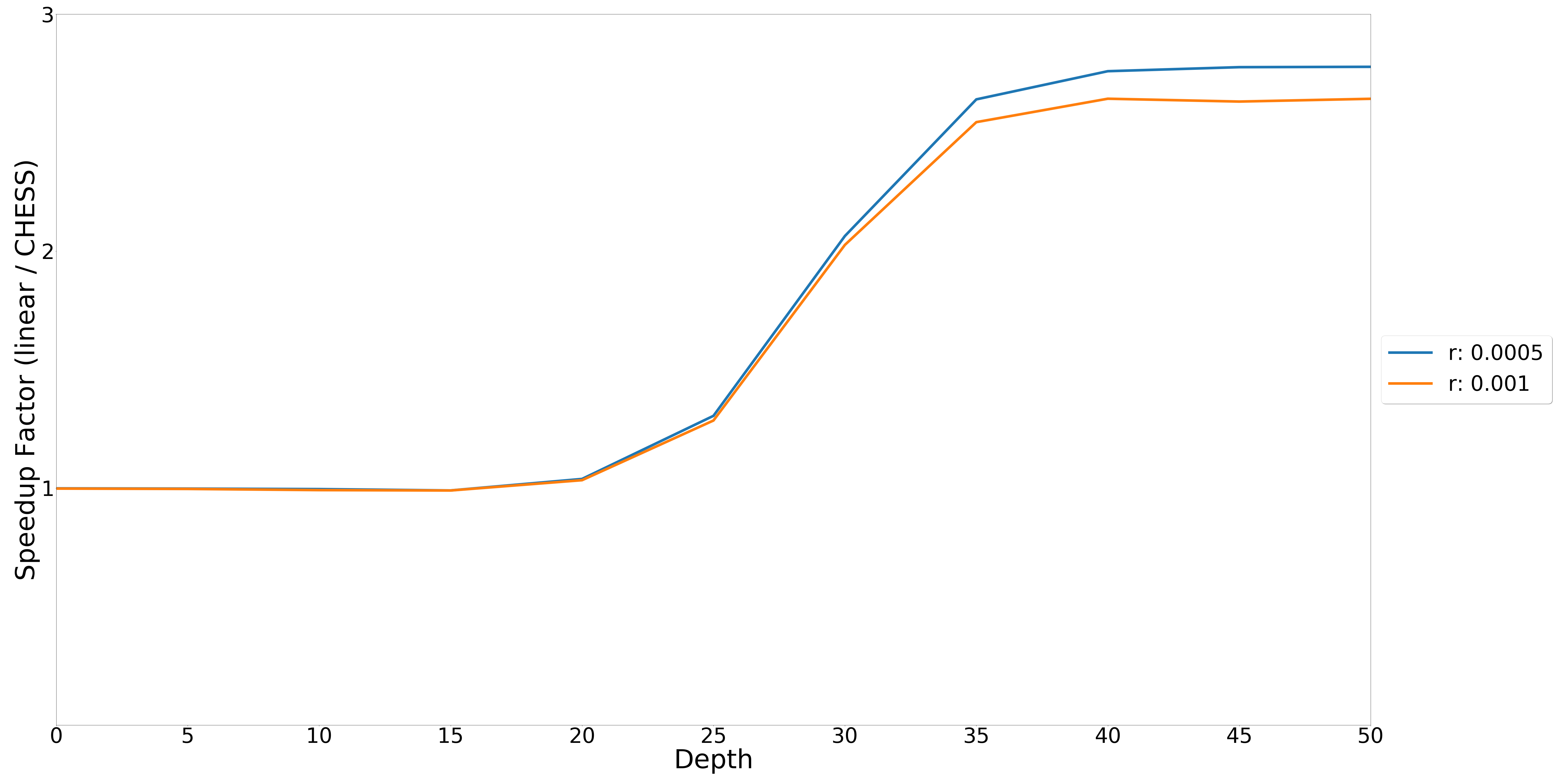}
\caption{Speedup Factor of CHESS on APOGEE vs. Na\"ive Search Under Cosine Distance as a Function of Search Depth}
\label{speedup_factor_cos_png}
\end{figure}

\begin{figure}[!t]
\centering
\includegraphics[width=3in]{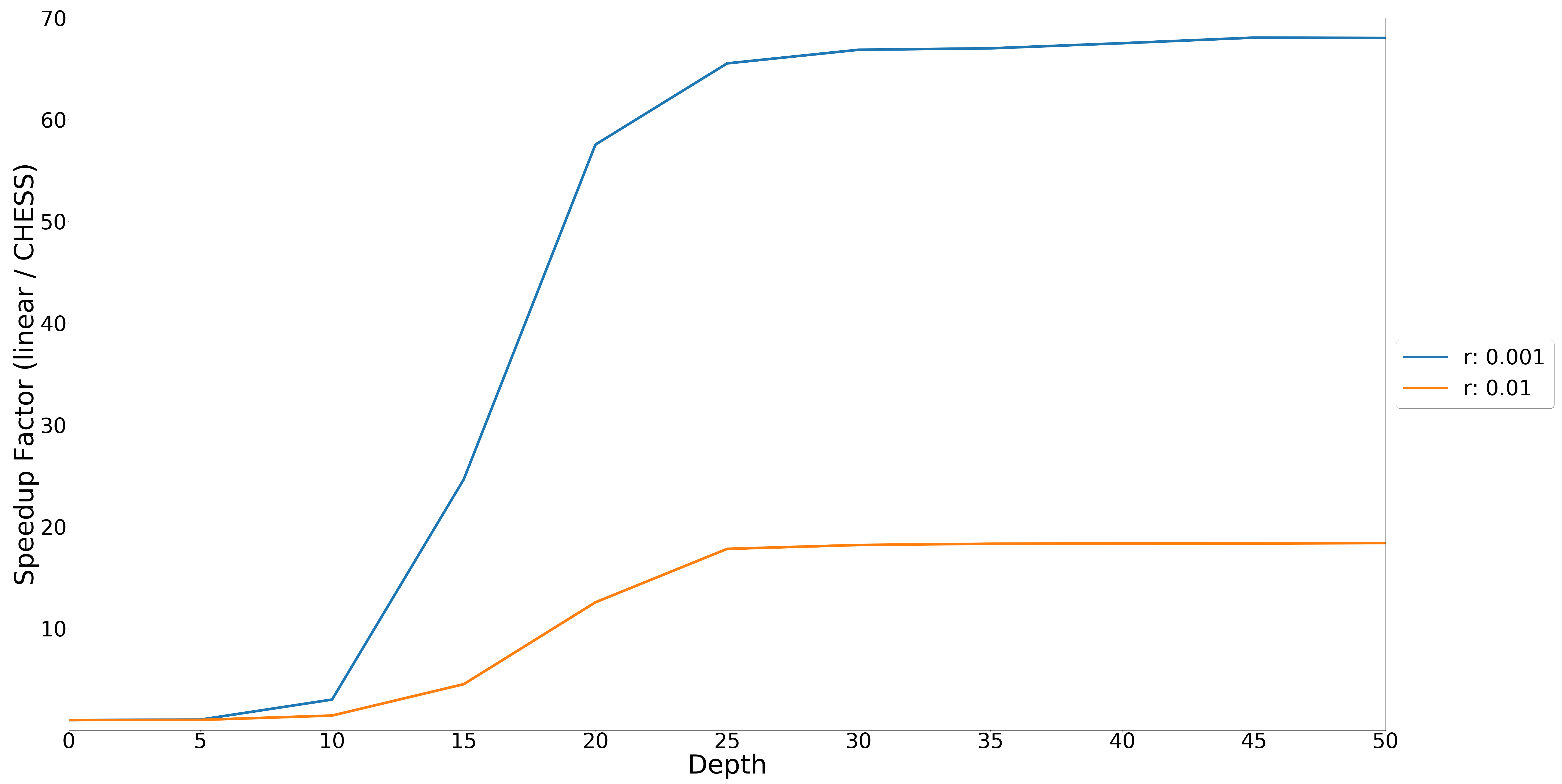}
\caption{Speedup Factor of CHESS on GreenGenes vs. Na\"ive Search Under Hamming Distance as a Function of Search Depth.}
\label{speedup_factor_hamming_png}
\end{figure}

On the GreenGenes data set, using Hamming distance, there is a clear reduction in search time as long as the search radius is small. Table~\ref{hamming} shows the number of comparisons, search time in seconds, the fraction of leaf clusters that must be searched, and the speedup factor vs. na\"ive linear search. At 99.9\% and 99\% sequence identity, speedup factors of 68.02 and 18.39 are observed. Figure~\ref{speedup_factor_hamming_png} illustrates that, for small search radii, CHESS search performance improves with depth. These small radii are biologically meaningful, as every sequence represents the same functional and structural component of the ribosome and, thus, almost the entire data set exists within a diameter of 90\% sequence identity.

\begin{table}[!t]
\renewcommand{\arraystretch}{1.3}
\caption{CHESS performance on GreenGenes vs. na\"ive, Hamming Distance}
\label{hamming}
\centering
\begin{tabular}{|c|c|c|c|c|c|c|}
\hline
\bfseries Depth & \multicolumn{2}{c|}{\textbf{Comparisons}} & \multicolumn{2}{c|}{\textbf{Search}} & {\textbf{Fraction}} & \bfseries Speedup\\
 & \multicolumn{2}{c|}{$\times 10^{5}$} & \multicolumn{2}{c|}{\textbf{Time (s)}} & {\textbf{Searched}} & \bfseries Factor\\
\hline
 & \bfseries $\mu$ & \bfseries $\sigma$ & \bfseries $\mu$ & \bfseries $\sigma$ & \bfseries $\mu$ & \bfseries $\mu$\\
\hline
\multicolumn{3}{|l}{\textbf{99.9\% Seq. Identity}} & \multicolumn{4}{|l|}{\textbf{Output Size: $\mu=245~  \sigma=486$}}\\
\hline
10 & 46.84 & 18.58 & 26.86 & 10.67 & 0.582 & 3.02\\
20 & 3.39 & 4.24 & 2.88 & 2.90 & 0.042 & 57.54\\
30 & 0.32 & 0.34 & 1.40 & 1.02 & 0.004 & 66.86\\
40 & 0.22 & 0.19 & 1.37 & 0.98 & 0.003 & 67.51\\
50 & 0.21 & 0.17 & 1.36 & 0.99 & 0.003 & 68.02\\
\hline
\multicolumn{3}{|l}{\textbf{99\% Seq. Identity}} & \multicolumn{4}{|l|}{\textbf{Output Size: $\mu=1,924~ \sigma=2,364$}}\\
\hline
10 & 59.09 & 14.46 & 33.86 & 8.21 & 0.734 & 1.45\\
20 & 12.59 & 9.05 & 9.82 & 6.45 & 0.156 & 12.57\\
30 & 2.79 & 2.82 & 5.77 & 4.22 & 0.035 & 18.20\\
40 & 2.66 & 2.69 & 5.58 & 3.98 & 0.033 & 18.34\\
50 & 2.66 & 2.69 & 5.57 & 3.98 & 0.033 & 18.39\\
\hline
\end{tabular}
\end{table}

FALCONN's benchmark results on APOGEE appear in Table~\ref{falconn-apogee}. While FALCONN is uniformly faster in terms of wall-clock time than CHESS, we note that the number of comparisons required by FALCONN is significantly greater than that required by CHESS. In particular, at a radius of 2,000 and a depth of 50, CHESS under L2 Norm requires, on average, only $6.8 \cdot 10^3$ distance comparisons; and at a radius of 0.0005 and a depth of 50, CHESS under Cosine Distance requires, on average, only $6.4 \cdot 10^3$ distance comparisons. FALCONN requires $1.32 \cdot 10^5$ comparisons in all cases, showing that CHESS provides an order of magnitude algorithmic improvement over FALCONN. In all cases, na\"ive search, CHESS, and FALCONN agreed on all search results, so we can consider all searches to be exact.

\begin{table}[!t]
\renewcommand{\arraystretch}{1.3}
\caption{FALCONN Search Performance on APOGEE}
\label{falconn-apogee}
\centering
\begin{tabular}{|c|c|c|c|c|}
\hline
\bfseries Distance & \bfseries Radius & \textbf{Comparisons} & \multicolumn{2}{c|}{\textbf{Search Time (s)}} \\
\hline
 &  &  $\times 10^5$ & \bfseries $\mu$ & \bfseries $\sigma$ \\

\hline
L2 & 2000 & 1.32 & 0.563 & 0.008 \\
 & 4000 & 1.32 & 0.564 & 0.008 \\
\hline
Cosine & 0.0005 & 1.32 & 0.520 & 0.014 \\
 & 0.001 & 1.32 & 0.517 & 0.011 \\
\hline
\end{tabular}
\end{table}

\subsection{Accuracy}\label{accuracy}

Similar to the argument presented in~\cite{yu2015quality}, the triangle inequality guarantees perfect accuracy as long as the distance function used is a metric. To search for all points within radius $r$ of a query, at each node we only need to look inside child-clusters whose centers are within radius $r+r_{i}$ of a query, where $r_i$ is the cluster radius of a child of that node.
False positives as compared against na\"ive search are not possible with CHESS. 
Under Hamming and Euclidean distances, which both obey the triangle inequality, the false negative rate is zero. Under cosine distance, which does not obey the triangle inequality, the false negative rate averages $4.4\times10^{-4}$ and is only nonzero at clustering depths greater than 30.

\subsection{Compression}

Our implementation of data compression as described in Section~\ref{compression} results in shrinking the APOGEE data set from 4 GB of storage (already represented as a NumPy memmap, a compact binary file) to 3.4 GB. With a larger data set and thus greater redundancy, we would expect greater compression. It may be worthwhile future work to investigate representing differences at each level of the binary tree, particularly for string or genomic data, where all differences are discrete.

\subsection{k-Nearest Neighbors Search Extension}

Machine learning and data science practitioners consider k-nearest neighbors (KNN) a nearly ubiquitous classification tool~\cite{muja2009fast, houle2005fast, hajebi2011fast, muja2014scalable}. While in this paper we have focused on the related $\rho$-nearest neighbors problem, it is worth discussing an extension to KNN. Under clustered search, KNN can be implemented as follows:

\begin{itemize}
    \item[]{Perform CHESS (Algorithm \ref{search}) with $\rho=\hat{r_c}$.}
    \item[]{If $|B_D(q, \rho)|>k$ and more than one leaf cluster was searched:}
        \subitem{Repeat CHESS, halving $\rho$ until $|B_D(q, \rho)|\leq k$.}
        \subitem{Double $\rho$ once.}
    \item[]{If $|B_D(q, \rho))<k$ and $\rho\leqslant r_{c_{0}}$:}
        \subitem{Repeat CHESS, doubling $\rho$ until $|B_D(q, \rho)|\geqslant k$.}
    \item[]{Return k-nearest neighbors from $B_D(q, \rho)$.}
\end{itemize}

where $\hat{r_c}$ is the \textit{median} cluster radius of leaf clusters, and $r_{c_{0}}$ is the radius of the root cluster.

The loops inside both conditionals are bound by $\log(r_{c_0})$ calls to the search function. The return statement has an asymptotic complexity of $O(|B_{D}(q, \rho)|\cdot \log{k})$ if a k-element min-heap is used to filter the results. Therefore the complexity of KNN search is:

\begin{gather}
O(S\cdot \log(r^\prime) + |B_D(q, \rho)|\cdot \log{k}) 
\label{knn-complexity}
\end{gather}
where $S$ is the asymptotic complexity of CHESS given in Equation~\ref{hierarchical-complexity} and the radius $r^\prime$ is $r_{c_0}$ when $r_{c_0}$ includes values greater than 1, and otherwise $\frac{1}{r_{c_0}}$.

\section{Conclusion and Discussion}

We have presented CHESS (Clustered Hierarchical Entropy-Scaling Search), an algorithm and software implementation for $\rho$-nearest neighbors approximate search of large data sets, which improves the coarse search time of entropy-scaling search~\cite{yu2015entropy} from $k$ to $log_2 k$, where $k$ is the number of clusters. Like \cite{yu2015entropy}, CHESS's asymptotic complexity is not defined in terms of $n$ (the size of the data set being searched. However, the $B_D(q,r)$ term (which is equivalent to the output size) has an \textit{implicit} dependence on the size of the data set; as more data are collected, particularly if they are similar to existing data, they are expected to increase the density of each cluster. This can be addressed by further deepening the hierarchical clustering, but this also creates an implicit dependence on $n$ (as $n$ grows, so $k$ should grow, but only as $\log(n)$).

CHESS is most effective when data exhibit low metric entropy and fractal dimension. CHESS requires an order of magnitude fewer comparisons than the FALCONN locality-sensitive hashing library, and provides data compression. While the wall-clock benchmarks slightly favor FALCONN, it is important to note that CHESS's Python implementation imposes some overhead, despite using the \texttt{sklearn} library for distance computations. In contrast, FALCONN relies on the highly-optimized \texttt{Eigen} linear-algebra library, which uses AVX intrinsics to achieve vectorization on the CPU. In addition, CHESS's I/O routines are in Python, which also adds overhead. A native implementation of CHESS in C++ or Rust will be worthwhile future engineering work. Unlike locality-sensitive hashing approaches, CHESS is extensible to any user-defined edit distance or similarity function. Examples include Jaccard similarity for sets, Wasserstein (Earth-Mover) distance for distributions, a variety of inter-graph distances for graphs, and others. As long as the distance function obeys the triangle inequality, search is guaranteed to be exact; non-metric distance functions such as cosine distance or BLAST E-values~\cite{altschul1990basic} can lead to false negatives, though not false positives. This is further discussed in section \ref{accuracy}.

CHESS did not begin to provide benefits on APOGEE with cosine distance until significantly higher cluster depths compared to L2 norm. This is at odds with the flat entropy-scaling search results from \cite{yu2015entropy}, though those were from a protein structure data set. Further investigation into cosine distance on APOGEE is warranted, but so is the investigation of other distance functions, such as Wasserstein or Earth Mover distance.

We have also discussed the theoretical properties of implementing k-nearest neighbor search using CHESS (as well as flat entropy-scaling search). The implementation is a topic for future work.

In our comparison to FALCONN~\cite{razenshteyn2015falconn}, it is worth noting that CHESS can be thought of as a form of locality-sensitive hashing~\cite{gionis1999similarity, har2012approximate}, though one not based on the choice of random separating hyperplanes. Similar objects hash, or cluster, into like clusters, but (other than the possibility of near neighbors appearing in neighboring clusters) it lacks the probabilistic nature of classical locality-sensitive hashing approaches. It is also reasonable to think of CHESS as a form of database index; if one wished to dispense with the data compression aspect, the cluster hierarchy itself could be thought of as an index. Indeed, we do not foresee any serious theoretical obstacles to incorporating such an index into a relational or key-value database such as PostgreSQL or MongoDB.

CHESS can easily be extended to use GPU-accelerated distance calculations. We provide some distance functions implemented in tensorflow. We did not use them for benchmarks because the overhead of sending data to the GPU is far greater than the computational cost of the distance functions used in this manuscript. For more costly distance functions, for example maximal common subgraph, GPU implementations may be worthwhile.

CHESS can easily be adapted to allow for live updates to the data. As new data points become available, we can perform a binary search with zero search radius to identify the leaf cluster to which the new point would best belong. If a point is added that is far outside a leaf cluster (for example, twice the cluster radius or more), a new cluster is created.

The original entropy-scaling search manuscript~\cite{yu2015entropy} led to a thoughtful discussion of the fundamental limits of search~\cite{kannan2015fundamental}. In this study, we have suggested a further lowering of these limits to be logarithmic rather than linear in the number of leaf clusters.

The source code for CHESS is available under an MIT license at \url{https://github.com/nishaq503/CHESS}.


\ifCLASSOPTIONcompsoc
  \section*{Acknowledgments}
\else
  \section*{Acknowledgment}
\fi

The authors would like to thank Bonnie Berger, Michael Baym, and Y. William Yu for suggesting the extension of entropy-scaling search to a hierarchical paradigm, and Tom Howard and Matthew Daily for helpful discussions.



\bibliographystyle{IEEEtran}
\bibliography{main.bib}
\end{document}